\documentclass[
	aps, pra, superscriptaddress, twocolumn,
	10pt
	floatfix, 
    nofootinbib,
	tightenlines
]{revtex4-1}
\usepackage[final]{graphicx}
\usepackage{times,bbm,amsmath,amssymb}
\usepackage{epsfig,color}
\usepackage{xcolor}
\usepackage{hyperref}
\hypersetup{
    colorlinks = true
}
\usepackage{cleveref}
\usepackage{microtype}

\usepackage{float,siunitx}
\usepackage[caption = false]{subfig}

\usepackage[greek,english]{babel}
\usepackage{thumbpdf,enumerate}
\usepackage{booktabs}
\usepackage{sidecap}
\usepackage[scaled=.8]{couriers}
\usepackage{multirow}
\usepackage{placeins}
\usepackage{relsize}
\usepackage{pst-grad,bm}
\usepackage{epigraph}
\usepackage{gensymb}
\usepackage{longtable}
\usepackage{ulem} 
\usepackage{acronym}
\usepackage{physics}
\usepackage{easyReview}

\newacro{ANN}{Artificial Neural Network}
\newacro{BS}{Beam Splitter}
\newacro{CCD}{Charge-Coupled Device}
\newacro{CNN}{Convolutional Neural Network}
\newacro{DoF}{Degree of Freedom}
\newacro{DQWL}{Discrete-time Quantum Walk on a Line}
\newacro{DR}{Dimensionality Reduction}
\newacro{ETC}{Extra Trees Classifier}
\newacro{ETR}{Extra Trees Regressor}
\newacro{FFT}{Fast Fourier Transform}
\newacro{GGM}{generalized Gell-Mann matrices}
\newacro{HG}{Hermite-Gauss}
\newacro{HWP}{Half-Wave Plate}
\newacro{HyGG}{Hypergeometric-Gaussian}
\newacro{K-NN}{K-nearest neighbour}
\newacro{LDA}{Linear Discriminant Analysis}
\newacro{LG}{Laguerre-Gauss}
\newacro{LLE}{Locally Linear Embedding}
\newacro{MDL}{minimum description length}
\newacro{ML}{Machine Learning}
\newacro{OAM}{Orbital Angular Momentum}
\newacro{PBS}{Polarizing Beam Splitter}
\newacro{PCA}{Principal Component Analysis}
\newacro{RBF}{Radial Basis Function}
\newacro{QP}{Q-Plate}
\newacro{QW}{Quantum Walk}
\newacro{QWP}{Quarter-Wave Plate}
\newacro{RFC}{Random Forest Classifier}
\newacro{RFR}{Random Forest Regressor}
\newacro{SAM}{Spin Angular Momentum}
\newacro{SLM}{Spatial Light Modulator}
\newacro{SMF}{Single-Mode Fiber}
\newacro{SVM}{Support Vector Machine}
\newacro{SVR}{Support Vector Regressor}
\newacro{TEM}{Transverse Electromagnetic}
\newacro{VVB}{Vector Vortex Beam}
\begin{document}


\address{Dipartimento di Fisica, Sapienza Universit\`{a} di Roma, Piazzale Aldo Moro 5, I-00185 Roma, Italy}




\vspace{10pt}

\title{Regression of high dimensional angular momentum  states of light} 

\author{Danilo Zia}

\author{Riccardo Checchinato}

\author{Alessia Suprano}

\author{ Taira Giordani}

\author{Emanuele Polino}
\address{Dipartimento di Fisica, Sapienza Universit\`{a} di Roma, Piazzale Aldo Moro 5, I-00185 Roma, Italy}
\author{Luca  Innocenti} 
\address{Centre for Theoretical Atomic, Molecular, and Optical Physics,
School of Mathematics and Physics, Queen's University Belfast, BT7 1NN Belfast, United Kingdom}
\address{Università degli Studi di Palermo, Dipartimento di Fisica e Chimica – Emilio Segrè, via Archirafi 36, I-90123 Palermo, Italy}
\author{Alessandro Ferraro} \author{ Mauro Paternostro} 

\address{Centre for Theoretical Atomic, Molecular, and Optical Physics,
School of Mathematics and Physics, Queen's University Belfast, BT7 1NN Belfast, United Kingdom}

\author{ Nicol\`o Spagnolo} 

\author{Fabio Sciarrino }

\email{fabio.sciarrino@uniroma1.it}

\address{Dipartimento di Fisica, Sapienza Universit\`{a} di Roma, Piazzale Aldo Moro 5, I-00185 Roma, Italy}

\begin{abstract}
The Orbital Angular Momentum (OAM) of light is an infinite-dimensional degree of freedom of light with several  applications in both classical and quantum optics.
However, to fully take advantage of the potential of OAM states, reliable detection platforms to characterize generated states in experimental conditions are needed.
Here, we present an approach to reconstruct input OAM states from measurements of the spatial intensity distributions they produce.
To obviate issues arising from intrinsic symmetry of Laguerre-Gauss modes, we employ a pair of intensity profiles per state projecting it only on two distinct bases, showing how this allows to uniquely recover input states from the collected data.
Our approach is based on a combined application of dimensionality reduction via principal component analysis, and linear regression, and thus has a low computational cost during both training and testing stages.
We showcase our approach in a real photonic setup, generating up-to-four-dimensional OAM states through a quantum walk dynamics.
The high performances and versatility of the demonstrated approach make it an ideal tool to characterize high-dimensional states in quantum information protocols.

\end{abstract}

\maketitle

\section{Introduction}
\label{sec:1}
The orbital angular momentum (OAM) of light is an internal degree of freedom associated to nontrivial transverse spatial wavefronts. As shown by Allen in his seminal paper~\cite{allen_0AM_1992}, helical beams with an azimuthal phase dependence $e^{i \ell \phi}$, with $\phi$ the azimuthal angle in cylindrical coordinates, carry an OAM equal to $\ell \hbar$ per photon. The OAM finds several optical applications in the classical regime, ranging from microparticle trapping~\cite{Zhan}  to communication~\cite{willner2015optical, bozinovic2013terabitscale, malik2012influence,baghdady2016multi,wang2016advances}.
Because the OAM can support countably infinitely many distinguishable states, it is ideally suitable to compress a large amount of information into single photons' states.
Being able to manipulate high-dimensional quantum states --- which we will refer to as \textit{qudits} in the following --- is highly beneficial for several quantum information protocols.
Such applications of OAM-based qudits include quantum simulation~\cite{cardano2015quantum,cardano2016statistical,cardano_zak_2017,Buluta2009}, metrology~\cite{fickler2012quantum,dambrosio_gear2013,cimini2021non,polino2020photonic} and communication~\cite{cozzolino2019air,Cozzolino_rev,Mirhosseini_2015,krenn2015twisted,Sit17,Cozzolino2019_fiber}.

However, the capability to accurately generate and detect OAM states remains a challenging task.
Detection techniques that have been proposed in the literature include interferometric schemes~\cite{leach2002measuring, sztul2006double,slussarenko2010polarizing,lavery2011robust}, the use of diffractive elements~\cite{berkhout2008method,ferreira2011fraunhofer,mazilu2012simultaneous,mourka2011visualization, hickmann2010unveiling}, tilted convex lens ~\cite{Vaity2013Measuring}, interference patterns with reference beams~\cite{d2017measuring, harris1994laser,padgett1996experiment, fu2020universal,ariyawansa2021amplitude}, methods exploiting Doppler frequency shift~\cite{courtial1998rotational, vasnetsov2003observation, zhou2017orbital}, weak measurements~\cite{malik2014direct}, metamaterials~\cite{karimi2009generation, karimi2012time, jin2016generation,devlin2017arbitrary,deng2018diatomic, guo2021spin, gao2018nonlinear,allen2019experimental} and  holographic techniques~\cite{mair2001entanglement,forbes2016creation,Qassim2014,giordani_2018,bouchard2018measuring,kaiser2009complete,schulze2013measurement,d2013test,pachava2019modal}.

Machine learning (ML) techniques have recently been shown as a valuable tool to overcome the many experimental and theoretical limitations related to reconstructing OAM states.
In particular, neural networks have been used to recognize and classify structured light states such as superposition of OAM~\cite{zhang2021recognition,doster2017machine, Park_18,na2021deep,wang2022deep,raskatla2022convolutional} and vector vortex beams~\cite{giordani2020machine,suprano2021enhanced}, also considering the propagation in turbulent environments~\cite{krenn_2014,krenn_2016,Lohani_18,liu2019deep,li2018joint, NEARY2020126058,Xie:15,lohani2018turbulence,bhusal2020spatial,zhan2021experimental, avramov2020machine,cox2022interferometric,teo2021benchmarking}. 
In this context, most of the efforts have been focused on detecting
the probability of finding OAM states in a fixed basis, as opposed to being able to resolve coherence terms between different modes.
However, the latter is of fundamental importance to completely reconstruct the state under analysis.
In particular, ML approaches can be used to reduce the number of measurements needed to recover the amplitudes and phases of the coefficients, making feasible the execution of quantum state tomography~\cite{torlai2018neural,rocchetto2018learning,carrasquilla2019reconstructing, lohani2021experimental, ma2021comparative, palmieri2020experimental, ahmed2021quantum}, a procedure that requires a number of measurements that scales quadratically with the state dimension \cite{thew2002qudit} unless we have prior information about the state \cite{gross2010quantum}. 

ML can also be used to directly recover the coefficients of a state in a given measurement basis. Convolutional neural networks (CNNs) have in particular been satisfactorily used for such regression task~\cite{silva2020machine}. However, the training of these CNN-based approaches involves in general a high computational cost. 

In this work, we present an approach to overcome these limitations thanks to the combined use of dimensionality reduction (DR)~\cite{abdi2010principal} and regression techniques. We find that, in particular, \textit{linear} regression approaches are best suited to solve the task at hand~\cite{quinlan1986induction, breiman2001random, scikit-learn}.
These algorithms allow us to deduce the Bloch vector of arbitrary superpositions of OAM states from the intensity profile obtained measuring them with a CCD camera, directly accessing the geometrical features of the states.
The improved computational efficiency of the methods we employ 
enables for a measurement scheme that can be more easily adapted to changes in the environmental conditions.

\begin{figure*}[t!]
    \centering
    \includegraphics[width=0.9\textwidth]{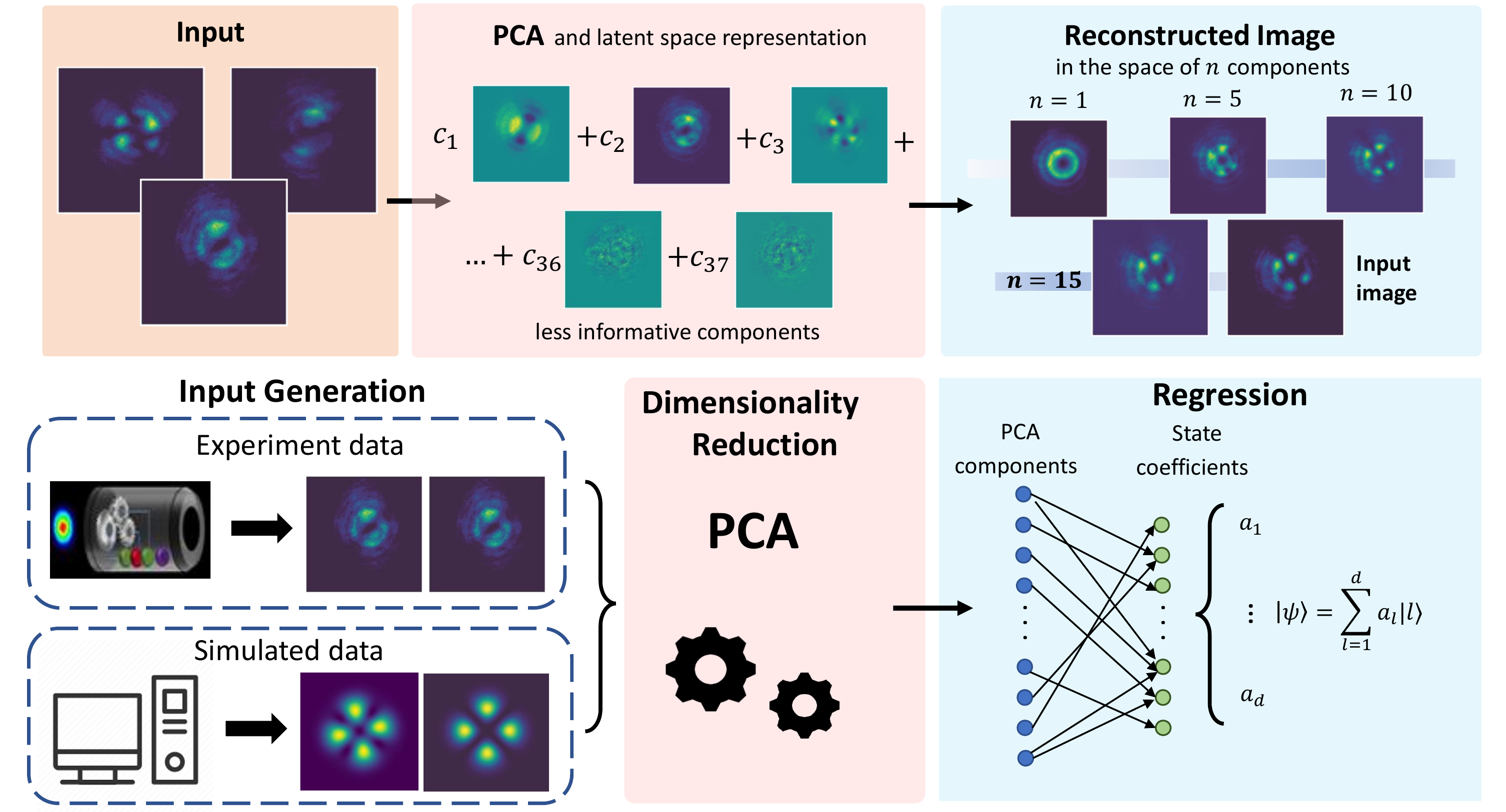}
    \caption{
    \textbf{Summary of the protocol.}
    (a) PCA is used to find a lower-dimensional representation of the
    64x64 pixel images.
    Working in an unsupervised fashion it derives the axes of the reduced space which best represent the input, in this specific case they are identified as images in which we can decompose the data.
    By tuning the number of reduced dimensions obtained via PCA, we have different reconstruction accuracies. 
    (b) The proposed method works through the following  3 steps. Firstly, the dataset is generated theoretically with a computer simulation or experimentally exploiting a setup suited for the engineering of OAM states. 
    Secondly, the data are given to the PCA algorithm that, reducing their dimensions, decreases the noise present in them and speeds up the training phase of the regressor. The latter, is finally used to obtain the coefficients of arbitrary superpositions of OAM modes. 
    }\label{fig:Schema_Con}
\end{figure*}

We firstly demonstrate the effectiveness of our approach on simulated data, obtained simulating the intensity profiles resulting from the measurement of generic OAM states with dimensions up to 8.
We then show the effectiveness of the approach in realistic experimental conditions, using it to characterise experimentally generated four-dimensional OAM states.
To generate the states, we employ a protocol based on quantum walk (QW) dynamics in the polarization and OAM degrees of freedom~ \cite{Innocenti2017,giordani_2018}, and measure the resulting states with CCD cameras.
An intrinsic issue of this type of measurement scheme is that it cannot directly distinguish between states with OAM numbers $\ket\ell$ and $\ket{-\ell}$.
We tackle this problem by performing, for each state, a measurement with \textit{two} CCD cameras, after splitting the incoming beam with a polarizing beamsplitter and using a q-plate (QP) on one arm in order to break the symmetry.
Compared to other methods proposed to break such degeneracy \cite{silva2020machine},
 our approach is fully generalizable and independent from the structure of the state. The states corresponding to simulated profiles can be perfectly reconstructed with a fidelity of $100\%$, while the application of such method on experimental states allows to reach average fidelities close to $97\%$. This demonstrates the power of our regression approach, and its promise to be a valuable tool for applications in quantum information science. The conceptual scheme describing the procedure followed in this work is reported in Fig. \ref{fig:Schema_Con}. 

\section{Machine learning for states regression}
\label{sec:3}%
OAM states
can be described via Laguerre-Gaussian (LG) modes, which form an orthonormal basis of eigenfunctions for the transverse spatial profile of light. More precisely, these are solutions of the paraxial Helmholtz equation, that can be expressed in cylindrical coordinates as
\begin{equation}\small
    \begin{aligned}
    LG_{p,l}&(\rho,\phi,\theta)  = \; \frac{C_{p,l}}{W(z)}\left(\frac{\sqrt{2}\;\rho}{W(z)}\right)^{\left|l\right|}
    \!\! L_{p}^{\left|l\right|}\left(\frac{2 \rho^2}{W^{2}(z)}\right)\\&\times \exp\left[-\frac{\rho^2}{W^{2}(z)}+il\phi -i\frac{k \rho z}{2\left(z^2+z_0^2\right)}+iN_{p,l}\zeta(z)\right]\!,\!
    \end{aligned}
    \label{laguerregauss}
\end{equation}
where $p,l$ are referred to as radial and azimuthal indices, and are related to the number of nodes of the transverse spatial profile and to the OAM eigenvalue, respectively. Furethermore, $L_{p}^{\left|l\right|}$ are the Laguerre polynomials, $W(z)$ is the beam waist after a propagation distance equal to $z$, $N_{p,l} = 2p+\left|l\right|+1$ is the mode order, $\zeta(z) = \arctan(2z/(kW(0)^2))$ is the Gouy phase and $C_{p,l}$ is a normalization constant. The exponential term $e^{i l\phi}$ characterizes the beam shape, giving it the well known helical structure. This phase term generates a singularity along the beam axis, so the intensity profiles associated with such modes present the peculiar doughnut shape. Throughout the work we consider only LG modes with radial index $p=0$. 

Our goal is to retrieve the complex amplitudes of given LG states with respect to the LG basis $\ket\ell$, from measurements of their intensity profiles.
To this end, we use a combination of dimensionality reduction and a regression algorithm.
Dimensionality reduction refers to a class of algorithms whose purpose is to find accurate lower-dimensional representations of high-dimensional data~\cite{fodor2002survey}.
More specifically, taking data in a high dimensional space $\mathbb{R}^{d}$, such as datasets of images with dimension equal to the number of pixels in the image, and mapping it into a new space $\mathbb{R}^{n}$ whose dimensionality $n$ is much smaller than the original one $d$.
Using this procedure to preprocess the dataset allows to reach a consistent speedup on the learning process performed by the regressor, thanks to the regression algorithm having to work on a much more compact representation of the data. 
More specifically, we use principal component analysis (PCA)~\cite{abdi2010principal} as dimensionality reduction algorithm.
PCA works by finding the linear subspace that optimally captures the variation in the input data, and thus provides a linear mapping from input data to a lower-dimensional \textit{latent space}, defined by the directions maximizing the variance of the projected data (see Fig. \ref{fig:Schema_Con}-a). As shown in Appendix \ref{appPCA}, the linear transformation provided by the PCA preserves the geometrical property of the state, allowing for a direct interpretation of the reduced dataset.
This also has an added advantage of providing methods more resilient to noise, due to dimensionality reduction looking for a representation of the data that best reflects the relevant features of the given data.
In fact the noise is mapped into the less representative dimensions that are cut during the reduction.
\begin{figure}[ht]
    \centering
    \includegraphics[width=0.85\columnwidth]{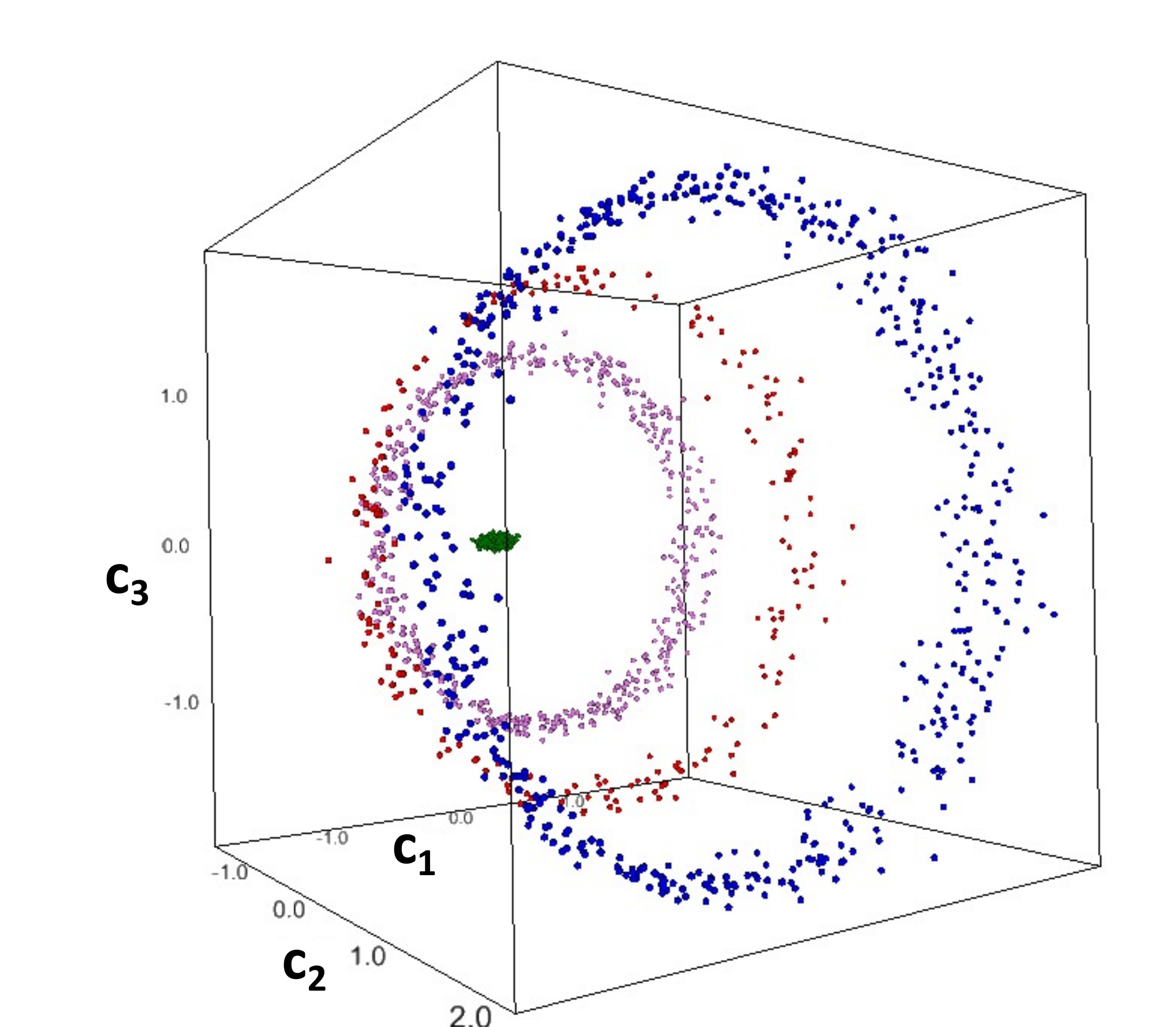}
    \caption{
    \textbf{Principal Component Analysis.}
    Representation of the simulated data reduced via PCA in the space composed by the first three components $\{c_{1}, c_{2}, c_{3} \}$ of the latent space. Here, we consider states of the form described by equation \ref{VVB}.
    Each color corresponds to a set of states corresponding to a fixed value of $\theta$. More specifically,
    $\theta=\pi$ (green), $\theta=7\pi/8$ (purple) $\theta=3\pi/4$ (red) and $\theta=\pi/2$ (blue). It can be noted that the circular structure given by the phase $\phi$ is preserved in latent space. This statement is also supported by the fact that the states with $\theta=\pi$, in green, are invariant with respect to the parameter $\phi$ and are mapped in the same region and not on a circle.
    }
    \label{fig:PCA_cerchi}
\end{figure}
Moreover, the application of \textit{linear} dimensionality reduction, in the form of PCA, is particularly suited to the task of reconstructing quantum states from measurement outcomes, due to the linearity intrinsic to this problem~\cite{giordani2020machine}.
To clarify this feature and the correlated advantages, consider states of the form
\begin{equation}\label{VVB}
    \vert \Psi \rangle =
    \cos{\frac{\theta}{2}}\ket1
    +e^{i\phi} \sin{\frac{\theta}{2}}\ket{-1},
\end{equation}
where $\theta \in [0,\pi]$ and $\phi \in [0,2\pi]$. Applying PCA on the training dataset, three dimensions of the latent space are sufficient to capture almost all of the relevant information.
We expect to retrieve in the latent space the Bloch sphere representation of the Hilbert space associated to one qubit. In fact, focusing only on four distinct classes characterized by $\theta=\pi, 7\pi/8, 3\pi/4, \pi/2$ and arbitrary $\phi$, the distribution of the dataset in the latent space is characterized by four circular structures with a growing radius that corresponds to the different $\theta$ values, while each of such circumferences is given by the parameter $\phi \in [0,2\pi]$ (see Fig. \ref{fig:PCA_cerchi}). Therefore, PCA preserves the geometrical feature of the space directly correlating the original parameters $\theta$ and $\phi$ with the position in the latent space. Such properties of the algorithm made it particularly suitable for preliminary data processing for both classification and regression tasks~\cite{giordani2020machine}.

Therefore, to retrieve the mapping between output intensity profiles and corresponding probabilities amplitudes, we train a linear regression algorithm on the data in the latent space obtained from PCA.
More specifically, we perform a supervised training of the regression using as objective the Bloch vector of the state under analysis. To address this problem in an high dimensional Hilbert space, we decompose the density matrix associated to the state using the Generalized Gellmann Matrices (GGM), a basis of orthogonal traceless operators, which can be used to define a Bloch representation for high-dimensional states.
To assess the accuracy of the results, we compute the fidelity between the Bloch vector predicted by the algorithm, and the corresponding true target state.
In order to avoid the possibility of non-physical state that might result from the algorithm, each state gets projected to the nearest pure physical state before calculating the fidelity.

The main problem 
for what concerns the regression of arbitrary OAM states using their intensity pattern is related to the following symmetry intrinsic to LG modes:
\begin{equation}\label{eq:symmetry}
    \left|\sum_{p,l}c_{p,l}LG_{p,l}\right|^2=\left|\sum_{p,l}c_{p,l}^*LG_{p,-l}\right|^2.
\end{equation}
This is due to LG modes having a dependence on $\ell$ such that $LG_{p,-\ell}=LG_{p,\ell}^*$.  
 A consequence of this symmetry is that different states might result in the same intensity profile, making it impossible to fully characterize input states from the acquired intensity profiles.
We addressed this problem both theoretically and experimentally acting on the modes for breaking this symmetry. 
In particular, our approach consists in a transformation of the reference state modifying the azimuthal index for all the modes which appear in the superposition. This produces two different superpositions and, hence, two distinct images that can be employed to reconstruct the state encoded in the initial image.
For example, let us consider a state $\ket{\Psi}$, defined as follows:
\begin{equation} \label{eq:general_symm1}
    \ket{\Psi} = a \ket{-2} + b \ket{-1} + c \ket{0} + d \ket{+1} + e \ket{+2}
\end{equation}
The symmetry rule in Eq.~\eqref{eq:symmetry} 
implies that the state $\ket{\Phi}$ which  is indistinguishable from $\ket{\Psi}$ 
according to the intensity profile measurements is:
\begin{equation} \label{eq:general_symm2}
    \ket{\Phi} = e^* \ket{-2} + d^* \ket{-1} + c^* \ket{0} + b^* \ket{+1} + a^* \ket{+2} 
\end{equation}
By increasing by one unit the OAM value of each mode in the superposition we obtain: 
\begin{equation} \label{eq:general_symm3}
    \begin{cases}
        \ket{\Psi'} = a \ket{-1} + b \ket{0} + c \ket{+1} + d \ket{+2} + e \ket{+3}\\
        \ket{\Phi'} = e^* \ket{-1} + d^* \ket{0} + c^* \ket{+1} + b^* \ket{+2} + a^* \ket{+3}
    \end{cases}
\end{equation}
The resulting states $\lbrace\ket{\Psi'},\ket{\Phi'} \rbrace$ are thus always distinguishable when $\ket\Psi$ and $\ket\Phi$ are not identical.

In other words, even though directly measuring the intensity profile of a given $\ket\Psi$ we cannot univocally determine that the input was $\ket\Psi$ rather than $\ket\Phi$, such degeneracy is lifted if for each state we measure both the intensity profile of $\ket\Psi$ and of the state obtained from $\ket\Psi$ after applying a transformation that increases each OAM values
by one unit.

\section{Results}
\label{sec:3}

In this section, we showcase the usefulness of our approach applying it to both simulated and experimental data.


\subsection{Numerical simulations}

We describe in this section the performances of our approach to retrieve a description of the quantum states corresponding to the observed intensity profiles, in a simulated regime.

The first scenario we consider is when the input states are superposition of only two orthogonal states, which can thus be described on a Bloch sphere.
In the simplest case, in which we only use one image for each state, the algorithm is unable to discern whether the state belongs to the left or the right hemisphere of the Bloch sphere.
This usually results in the model placing the states in the middle of the two hemispheres, along the equator. 
To avoid the degeneracy due to the symmetry inherent to OAM states, we use the protocol previously described, using a pair of intensity profiles per state.

\begin{figure}[!htb]
    \centering
    \includegraphics[width=1\columnwidth]{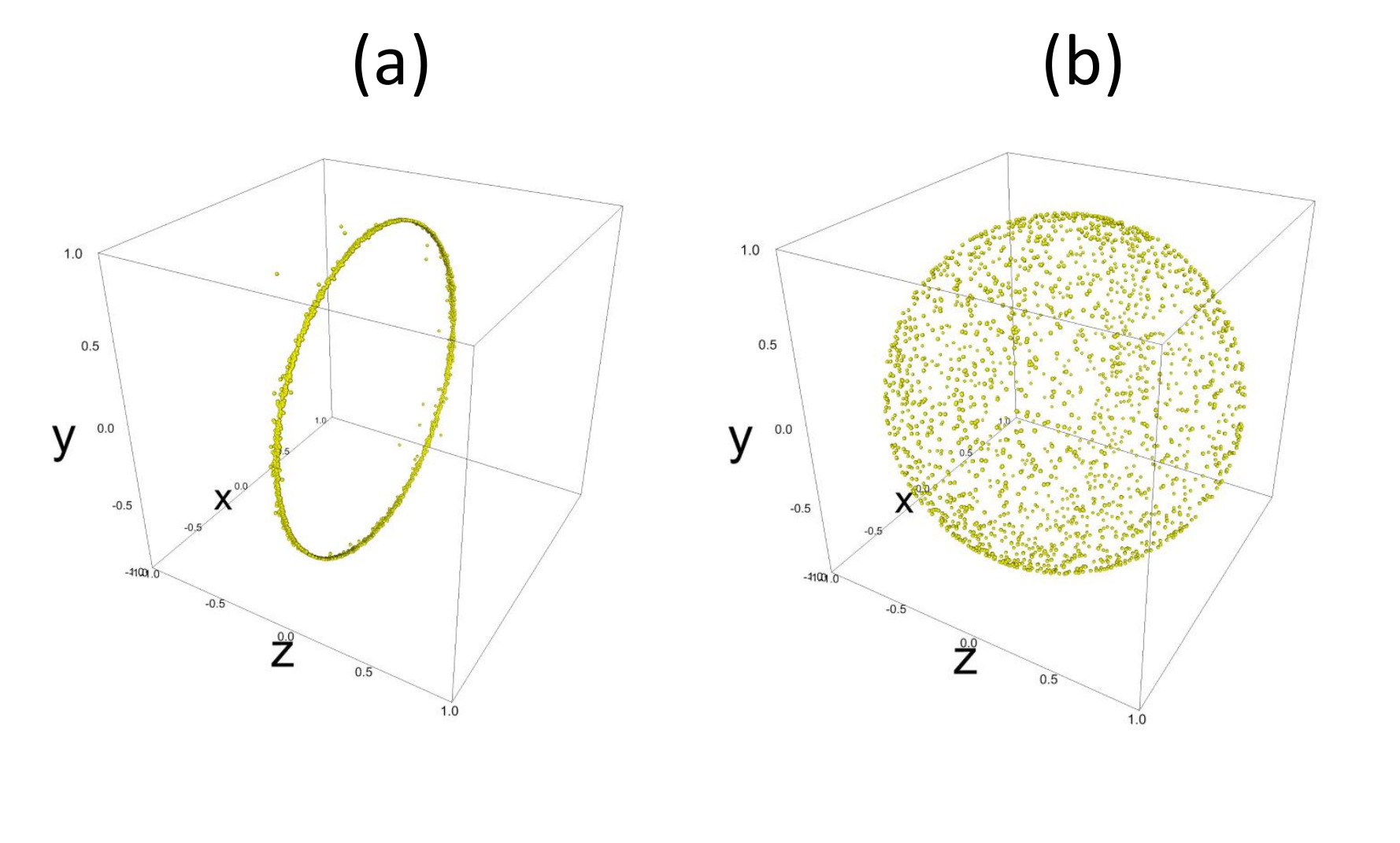}
    \caption{\textbf{Two-dimensional symmetry breaking.} Representation of the output of the regression algorithm on theoretical superpositions with $l \in \lbrace -1,+1 \rbrace$, after the projection on the nearest pure state. 
    We compare the position of each state on the Bloch sphere obtained from the regressor using  one image (a) and two images (b) per state.
    In panel (a) the higher values are obtained on the equator, when considering the poles to be along the $z$ axis, this is the strategy adopted by the regressor to minimize the errors. In fact, not being able to distinguish states placed on the two semispheres, this approach enables the regressor to obtain a higher mean fidelity. Instead in panel (b) the effects of symmetry breaking are evident, here the regressed states are placed near their real position.
    }
    \label{fig:PCA_1step_nosfera-tu}
\end{figure}

In this condition, thanks to the additional information provided by the images obtained augmenting the topological charge by one, the algorithm is capable of mapping the states on the spherical surface instead of accumulating them on the equator.
This behavior is explicitly shown in Fig. \ref{fig:PCA_1step_nosfera-tu}, in which the positions of the states in the Bloch space are reported for both the approaches. Similar results are also observed when the number of dimensions increases, more details on it and on the symmetry breaking can be found in Appendix \ref{App_Simmetria}.

\begin{figure*}[t!]
    \centering
    \includegraphics[width=1\textwidth]{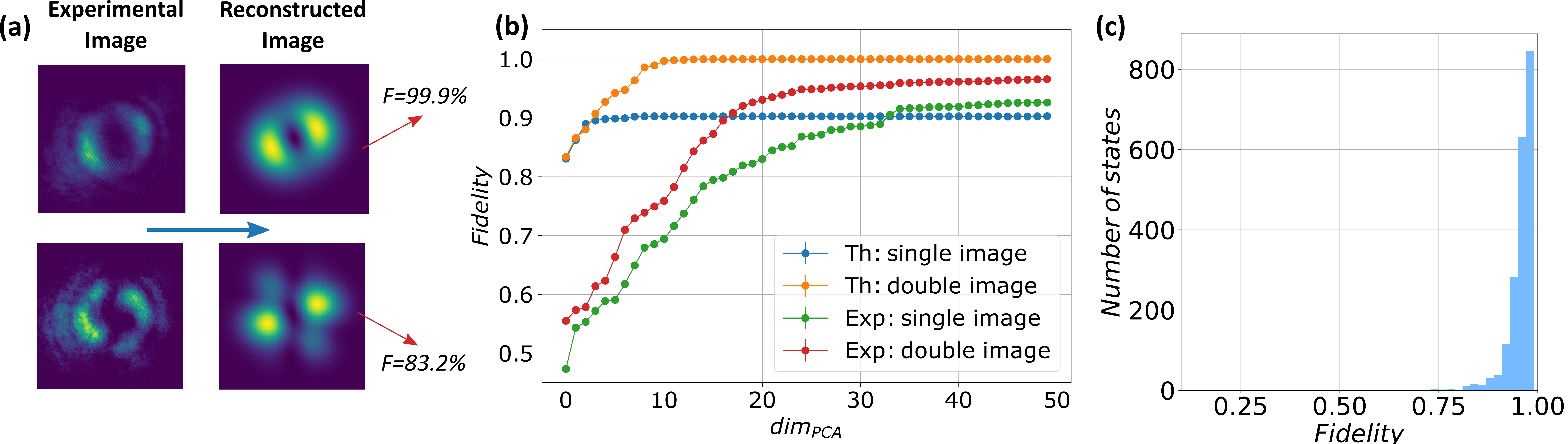}
    \caption{\textbf{Experimental Results.} (a) Comparison between the intensity distributions of collected experimental states and of reconstructed states whose coefficients of the LG superposition are retrieved through the regression analysis. To showcase the performances of the method, results for values of the fidelity equal to $ F = 99.9 \% $ and $ F = 83.2 \% $ are reported. (b) Trend of average fidelity obtained by increasing the number of PCA dimensions passed to the regression algorithm by state in the space spanned by $l \in \lbrace -3,-1,+1,+3 \rbrace$. Theoretically, the regression algorithm approach its maximum fidelity with PCA dimensions close to 15. Using single images does not allow to reach fidelity higher that $ F = 90 \% $ (blue line), while employing double images the fidelity value approach the $ 100 \% $ (orange line). Although experimental imperfections appear to break the symmetry feature also in the single image configuration (green line), the exploitation of double images increases the performances (red line). The points constituting all line are obtained averaging over $ 2000 $ random states, while the correspondent error is not appreciable with respect the size of the marker. (c) Experimental fidelity distributions calculated over 2000 random states in the double image configuration for state experimentally engineered in the space spanned by $l \in \lbrace -3,-1,+1,+3 \rbrace$. The mean value of the fidelity is $ F = 0.9661 \pm 0.0009$, where the error is given by the standard deviation on the average.
    }\label{fig:Exp_Res}
\end{figure*}

For our tests, we used random superpositions of states spanned by $d\in[2,8]$ orthogonal OAM basis states.
For any $d$, we simulate $10^4$ random pure states,
using $80\%$ to train the algorithm and the remaining $20\%$ as a test dataset.
For each state, we generate both the image associated to its intensity profile, and the image corresponding to the state obtained by increasing the values of the OAM to break the degeneracy.
Both images, composed of 64x64 pixels, are used as input of the PCA, and their resulting compressed representation is then fed to the linear regressor~\cite{scikit-learn} to solve the regression task.
To show the effectiveness of the procedure, we compared the results of the regression obtained using only the intensity profile of the state with those reached by processing the augmented dataset, containing also the profile of the superposition with increased azimuthal index. 

We show in Fig.~\ref{fig:Exp_Res}-b the behaviour of the fidelity as a function of the number of PCA dimensions in the $d=4$ case.
Using the information stored in the second image allows the regressor to achieve better performances in all cases considered.
In fact, we achieve fidelities above $90 \%$ considering only 5 dimensions for the reduced space, 
while a unit fidelity is reached using PCA to obtain a compressed representation in $d^2-1=15$ dimensions.
The plot ~\ref{fig:Exp_Res}-b also shows the effect of the symmetry breaking: in fact, using only one image per state, even in higher dimensional latent spaces, the performances are always significantly worst. 

Similar results are obtained for $d \in \{2,3,5,6,7,8\}$: the protocol always achieves unit fidelity applying linear regression on $d^2-1$ dimensions, and using the information stored in the second image (for more details see Appendix~\ref{Confronto}).

In conclusion, the additional information enables the regressor to solve the degenerancy and to make more accurate predictions. Moreover, using linear regression turns out to be optimal to directly connect the outputs of the PCA to the coefficients in the Bloch vector representation, allowing for better performances, compared to more complex regression algorithms such as the nonlinear Extra Tree Regressor (ETR). The latter leverages on the construction of a decision tree in which the branches are followed relying on the features of the input data, in this case the regressed Bloch vector is contained in the final leaf of the tree. In particular, the ETR reaches an accuracy equal to 100\% only in the bi-dimensional case, showing a damping in the performances when the state dimension increases (see Appendix~\ref{Confronto} for a comparison analysis). 

\subsection{Experimental implementation}
\begin{figure*}[ht!]
    \centering
    \includegraphics[width=0.9\textwidth]{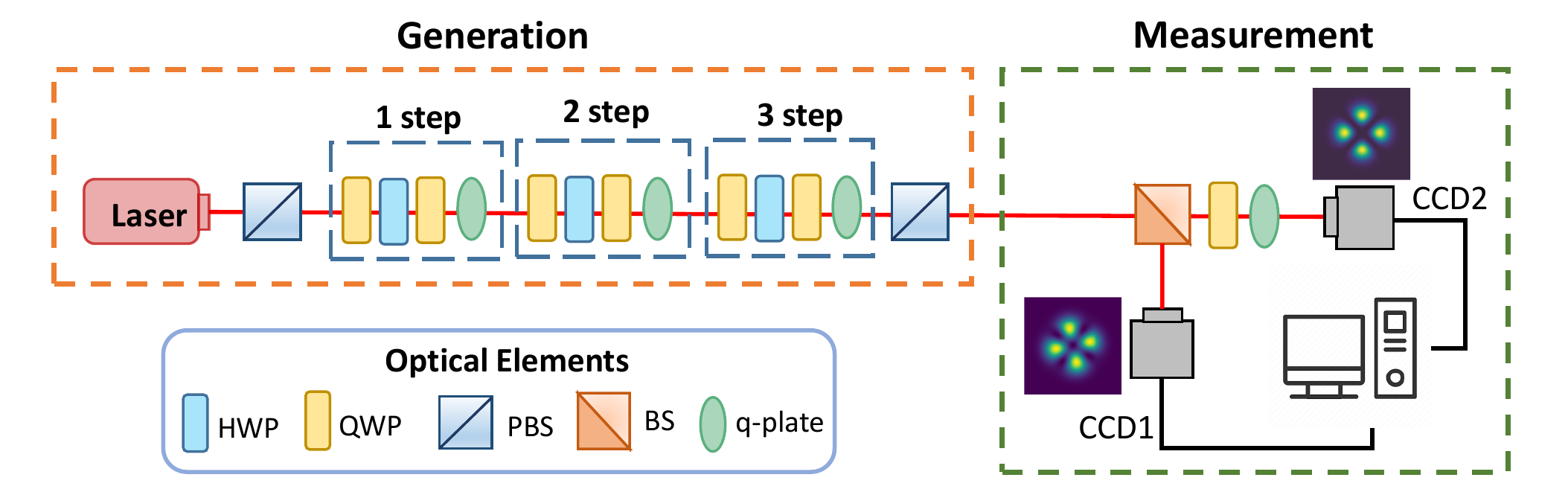}
    \caption{\textbf{Experimental Setup.} Diagram of the experimental apparatus used in the generation and measurement of spatially modulated photonic beams. The setup is composed by three blocks containing a cascade of waveplates, half-waveplate (HWP) and quarter-waveplates (QWP), followed by a q-plate (QP). Each block implements a step of the quantum walks dynamics. The input state is a Gaussian mode obtained thorught the coupling of a 808 nm laser to single mode fiber (SMF). After tracing out the polarization degree of freedom, the intensity profile corresponding to the resulting OAM state is recorded with a CCD camera.
    To uniquely retrieve input states from intensity profiles, the measurement stage uses two CCD cameras, one of which is placed after a QWP and a QP. This arrangement allows to break the symmetry inherent to LG modes.
    In this configuration the last waveplate is rotated at the angle $-\pi/4$ in order to change the horizontal polarization coming from the projection step, into the left circular one ($\ket{H}\rightarrow\ket{L}$). This allows the \ac{QP} to increase the \ac{OAM} of each mode by one.}
    \label{fig:Exp_setup}
\end{figure*}

To showcase the performance of the developed methodology we apply it to a real experimental scenario. Through a QW-based architecture~\cite{giordani_2018,Innocenti2017} we engineer OAM states of the form
\begin{equation} 
\ket{\psi} = a\ket{-3}+b\ket{-1}+c\ket{1}+d\ket{3},
\end{equation}
where $a,b,c,d\in\mathbb{C}$ can be chosen arbitrarily by tuning the parameters of the setup.

The implemented QW dynamics in a photonic platform exploits the two components of the photons angular momentum, the spin angular momentum and the OAM to encode respectively the coin and  walker states~\cite{giordani_2018}. By acting on the polarization degree of freedom it is possible to control the generated OAM states.

In particular, as shown in Fig.~\ref{fig:Exp_setup}, the setup is composed of three blocks containing a series of waveplates, acting on the coin state, followed by a q-plate. The latter is a device composed of a birefringent and inhomogeneous material capable of modify the photons' OAM conditionally on their polarization~\cite{marrucci-2006spin-to-orbital}, and is thus suitable to engineer nontrivial OAM states~\cite{giordani_2018,suprano2021dynamical}.

At the end of the QW, after a projection on the polarization space, the intensity distribution of the state is collected with CCD cameras. 
More specifically, the beam obtained at output of the QW is passed through a beam splitter (BS). We directly measure the beam with a CCD on one output arm (CCD1 in Fig.~\ref{fig:Exp_setup}), while on the other arm the measurement is performed after the evolution through an HWP and a QP, in order to increase each OAM value in the superposition by one unit and thus break the symmetry in~\eqref{eq:symmetry} (CCD2 in Fig.~\ref{fig:Exp_setup}).
We thus generate and measure $10^4$ random states.

The images collected by the CCDs are 1280x1024 pixels, but we scale them down to 64x64 pixels before feeding them to the algorithm.
We employ two different PCAs separately on the dataset. In particular, the first one is used to reduce the dimensions of the image of the states, while the second one is applied in the same manner to the superposition state with the augmented OAM values. This approach shows an increase in the performances of the method. 
The compressed representations of the images are then used to train the regressor.
Indeed, the two-image method allows us to reach a faster convergence to mean fidelities that are obtained by the one-image approach only when a large number of PCA dimensions is used.  
The results reported in Fig. \ref{fig:Exp_Res} are averaged on a test dataset composed by 2000 images for the case in which the training step is performed on 8000 samples and the regression algorithm is applied on the information stored in the first $50$ PCA dimensions.

These results showcase the high performances of our approach to characterize input states from measurement data in real experimental scenarios.

\section{Conclusions}
\label{sec:5}

We proposed and experimentally demonstrated a machine-learning-based approach to tackle the regression task of characterizing input OAM states from measurement outcomes.
We demonstrated our method for simulated states spanning up to eight dimensional spaces, showcasing its high performances even in noisy experimental conditions in a four dimensional space.
To solve the issue arising from the intrinsic symmetry of LG modes, we implemented a strategy to augment the number of intensities acquired per each state, thus allowing to unambiguously reconstruct input states from outcome intensity profiles.
From an experimental point of view, this approach is simply obtained through a beam splitter, and a set composed of a quarter-waveplate followed by a q-plate. This setup allows collecting two intensity distributions per state corresponding to projection on two distinct bases. Therefore, acting on the beam on a line without the need for interferometric or  holographic techniques, we are able to perform the required measurement of the intensity profile. This approach
is thus easily implementable, and effective to break the symmetry causing different OAMs to appear identical.
Moreover, the demonstrated protocol is not restricted by the dimension of the input states. Together with the feasibility of the experimental implementation, this make our approach an effective tool to characterize OAM states.
The fast training allowed by the use of PCA and linear regression makes for a highly versatile experimental detection scheme.
Although in this paper we used coherent states generated from a laser source, the same approach can be used at the single photon level using single-photon CCD cameras, and thus paving the way for ML-based quantum state tomography protocols.
Our protocol 
is thus promising for quantum technology applications that require the information encoded in OAM states and need fast and direct tracking. 


\section*{Acknowledgments}
We acknowledge support from the European Union's Horizon 2020 research and innovation programme (Future and Emerging Technologies) through project CANCER SCAN (grant nr. 828978).
L.I. acknowledges support from MUR and AWS under project 
PON Ricerca e Innovazione 2014-2020, ``calcolo quantistico in dispositivi quantistici rumorosi nel regime di scala intermedia" (NISQ - Noisy, Intermediate-Scale Quantum). MP acknowledges support by the European Union's Horizon 2020 FET-Open project TEQ (766900), the Leverhulme Trust Research Project Grant UltraQuTe (grant RGP-2018-266), the Royal Society Wolfson Fellowship (RSWF/R3/183013), the UK EPSRC (grant EP/T028424/1) and the Department for the Economy Northern Ireland under the US-Ireland R\&D Partnership Programme.

\appendix
\section{Linearity of PCA and phase mapping in the latent space}
\label{appPCA}

The principal component analysis is one of the most employed dimensionality reduction algorithms. Since the PCA is an example of unsupervised learning it does not have information about the data that are given to it other than the data itself. The operation done by PCA can be described as mapping the data into a linear combination along the most representative axes, the ones that maximize the variance among the data in the new space. \begin{figure}[!hbt!]
    \centering
    \includegraphics[width=0.75 \columnwidth]{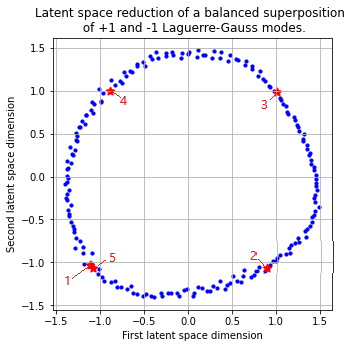}
    \includegraphics[width=0.48\textwidth]{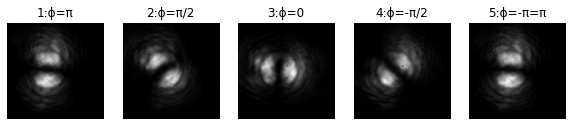}
    \caption{Latent space representation of the phase profiles of a simple experimental superposition of \ac{LG} modes $\ket{\psi}=1/\sqrt{2}\left(\ket{+1}+e^{i\phi}\ket{-1}\right)$. 
    The various points correspond to different values of the phase angle $\phi$. The five images at the bottom correspond to the red star symbols in the main graph at the top. It should be noted that the change in the phase angle $\phi$ generates a rotation in the respective images.
    This effect is then reflected in the angle at which each image can be found in the latent space representation.
    }
    \label{fig:PCA_cerchio}
\end{figure}

To clearly describe the process performed by the PCA, let us consider our data in the form of $m$ d-dimensional vectors $\lbrace x_1, x_2,\dots,x_m\rbrace$ in $\mathbb{R}^d$.
The aim is to reduce the dimensionality of these vectors using a linear transformation.
Firstly, a matrix $W \in \mathbb{R}^{n,d}$, with $n<d$, defines a map from the input data $x \in \mathbb{R}^{d}$ to a vector belonging to a lower-dimensional space $y \in \mathbb{R}^{n}$.
Then, it is possible to recover an approximation of the original vector $x$ from its compressed version.
More precisely, given the compressed vector $y=Wx$ in the low dimensional space $\mathbb{R}^n$, usually referred to as \emph{latent space}, it is possible to construct the recovered version $\Tilde{x}=Uy=UWx$ of the vector $x$, which also resides in the original high dimensional space $\mathbb{R}^d$.
In the PCA the compression $W$ and recovery $U$ matrices are obtained by minimizing the squared distance between the original $x$ and recovered $\Tilde{x}$ vectors.
Formally, we aim to solve the following minimization problem :
\begin{equation}\label{eq:PCA_minimization}
    \min_{W \in \mathbb{R}^{n,d}, U \in \mathbb{R}^{d,n}}
    \sum_{i=1}^{m} |x_i-UWx_i|^2.
\end{equation}

The linearity of the described mapping gives as a result that many intuitive properties are kept in the \emph{latent space}. For example, a shift in the phase of a state in a superposition of \ac{LG} modes is generally represented by a sort of rotation in the intensity profile, more evident when there are only two states.
As a result the images that show this type of difference are generally mapped to circularly shaped clusters in the latent space.
This gives, for example, easy access to information regarding the phase difference between modes directly from the latent space, since the linearity of the mapping preserves the inner structure of the data. 
This concept is illustrated in Fig.~\ref{fig:PCA_cerchio} where the position of different images in the latent space representation is compared to their contents. Therefore, this showcases how the latent space description is directly linked to the geometrical properties of the state.

\begin{figure}[!htb]
\centering
{Single Image Configuration}\\
\begin{minipage}[b]{0.48\columnwidth}
\subfloat[Correct coefficients]{\includegraphics[width=0.98\textwidth]{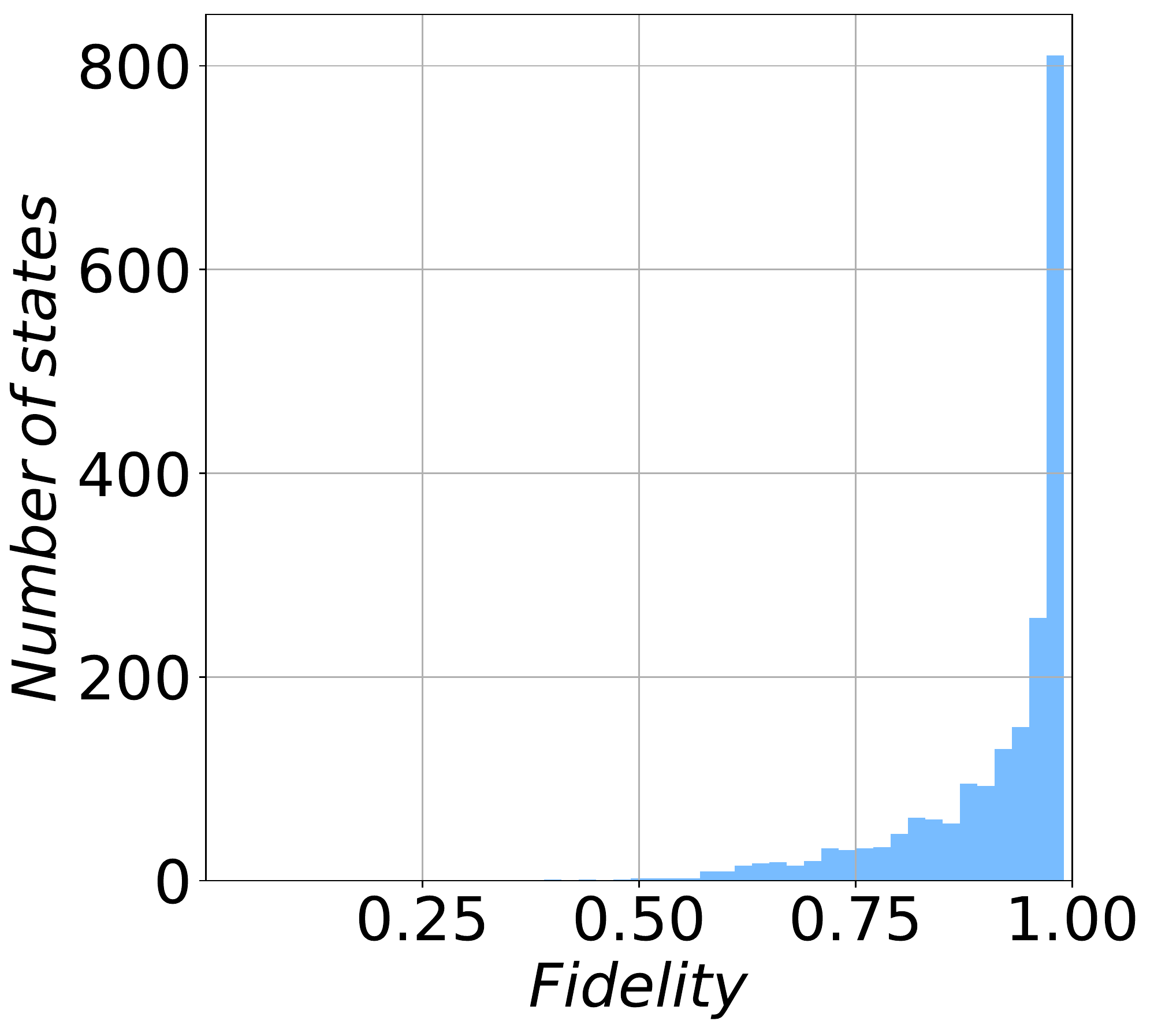}
    \label{fig:Fid_1-1_corr}}
\end{minipage}
\begin{minipage}[b]{0.48\columnwidth}
\centering
\subfloat[Inverted coefficients]{\includegraphics[width=0.98\textwidth]{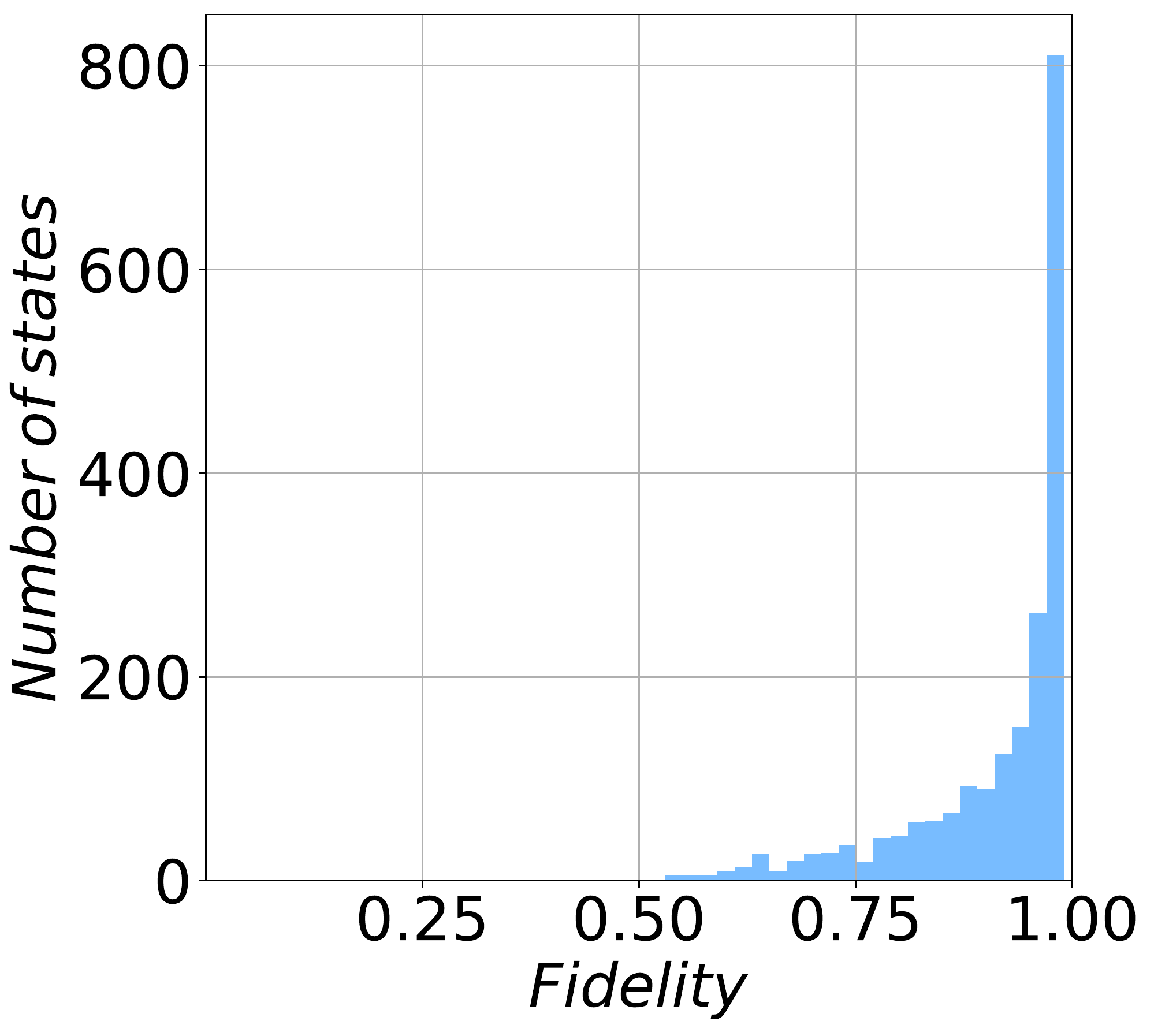}
    \label{fig:Fid_1-1_inv}}
\end{minipage}
\centering
\\
{Double Image Configuration}\\
\begin{minipage}[b]{0.48\columnwidth}
\subfloat[Correct coefficients]{\includegraphics[width=0.98\textwidth]{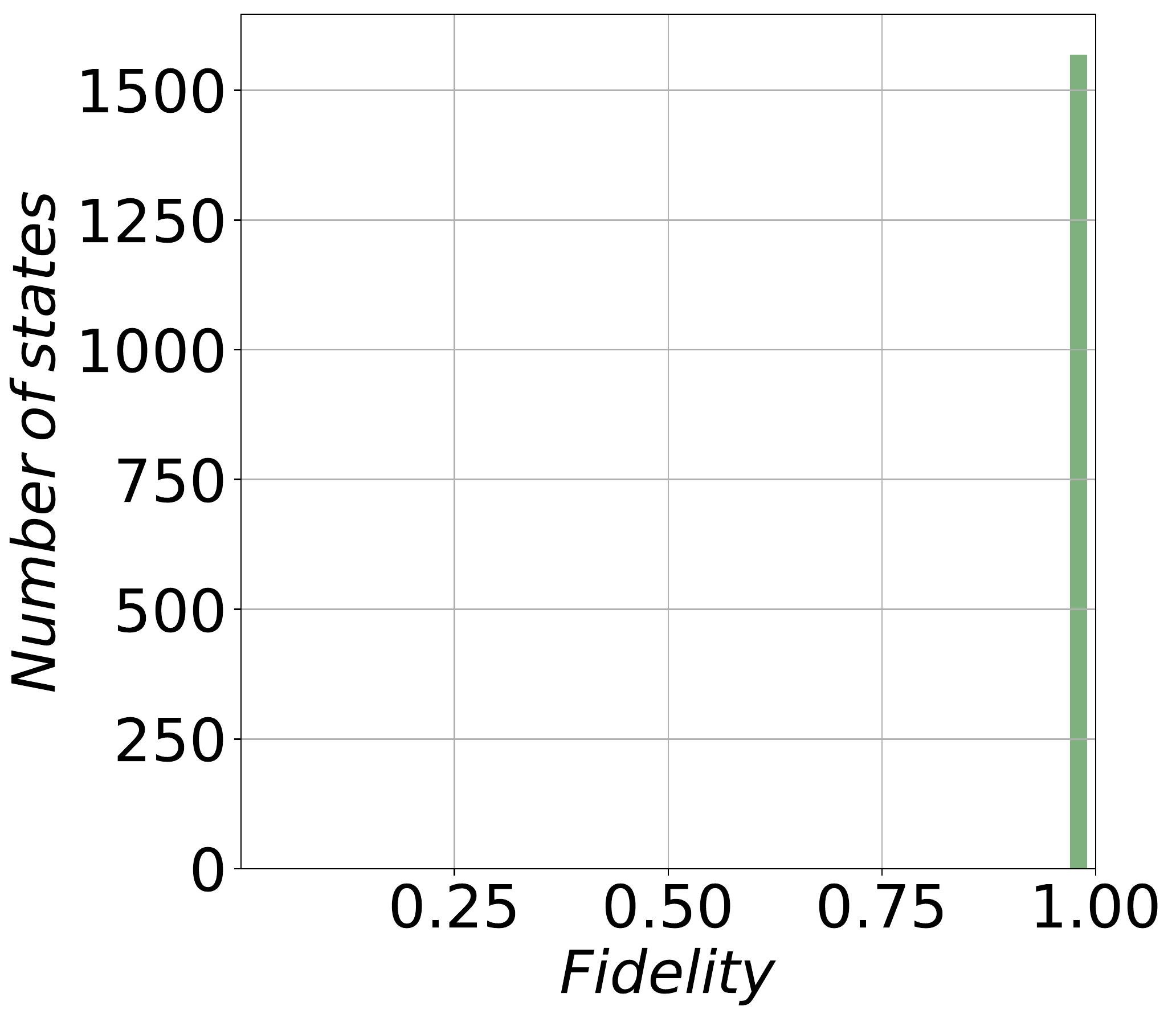}
    \label{fig:Fid_1-1_corr_double}}
\end{minipage}
\begin{minipage}[b]{0.48\columnwidth}
\centering
\subfloat[Inverted coefficients]{\includegraphics[width=0.98\textwidth]{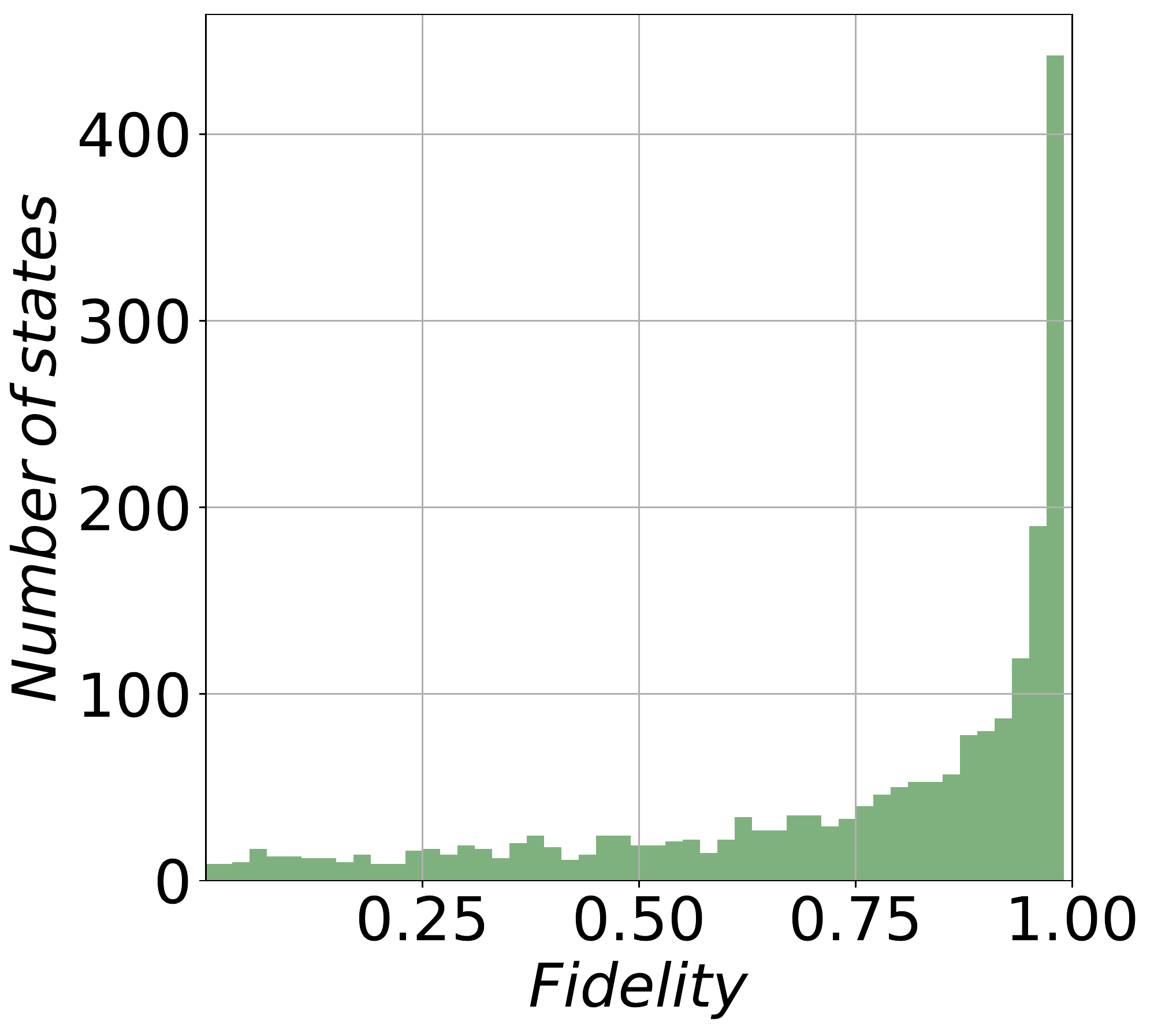}
    \label{fig:Fid_1-1_inv_double}}
\end{minipage}
\caption[.]{\textbf{Two-dimensional fidelities distribution.} Fidelity calculated on superpositions with $l \in \lbrace -1,+1 \rbrace$. Using only one image per state, the values of fidelity were calculated between the output of the regression with both the expected theoretical state (a) and the theoretical state after the inversion of the coefficients described by Eq.~\eqref{eq:symmetry} (b).
    The mean value of the fidelity in both graphs is $\Bar{F}_a=0.923(2)$ and $\Bar{F}_b=0.923(2)$, respectively. 
    They are compatible within the statistical error and thus the process can not distinguish the two cases.
    Computing the same fidelity in the double image configuration, the mean value of graph in panel (c) is $\Bar{F}_c=1$, while the mean value of graph in panel (d) is $\Bar{F}_{d}=0.764(6)$.
    They are incompatible and thus the process has broken the symmetry.}\label{fig:Fid_1-1_corretta_inversa}
\end{figure}

\begin{figure}[t!]
\centering
{Single Image Configuration}\\
\begin{minipage}[b]{0.48\columnwidth}
\subfloat[Correct coefficients]{\includegraphics[width=0.98\textwidth]{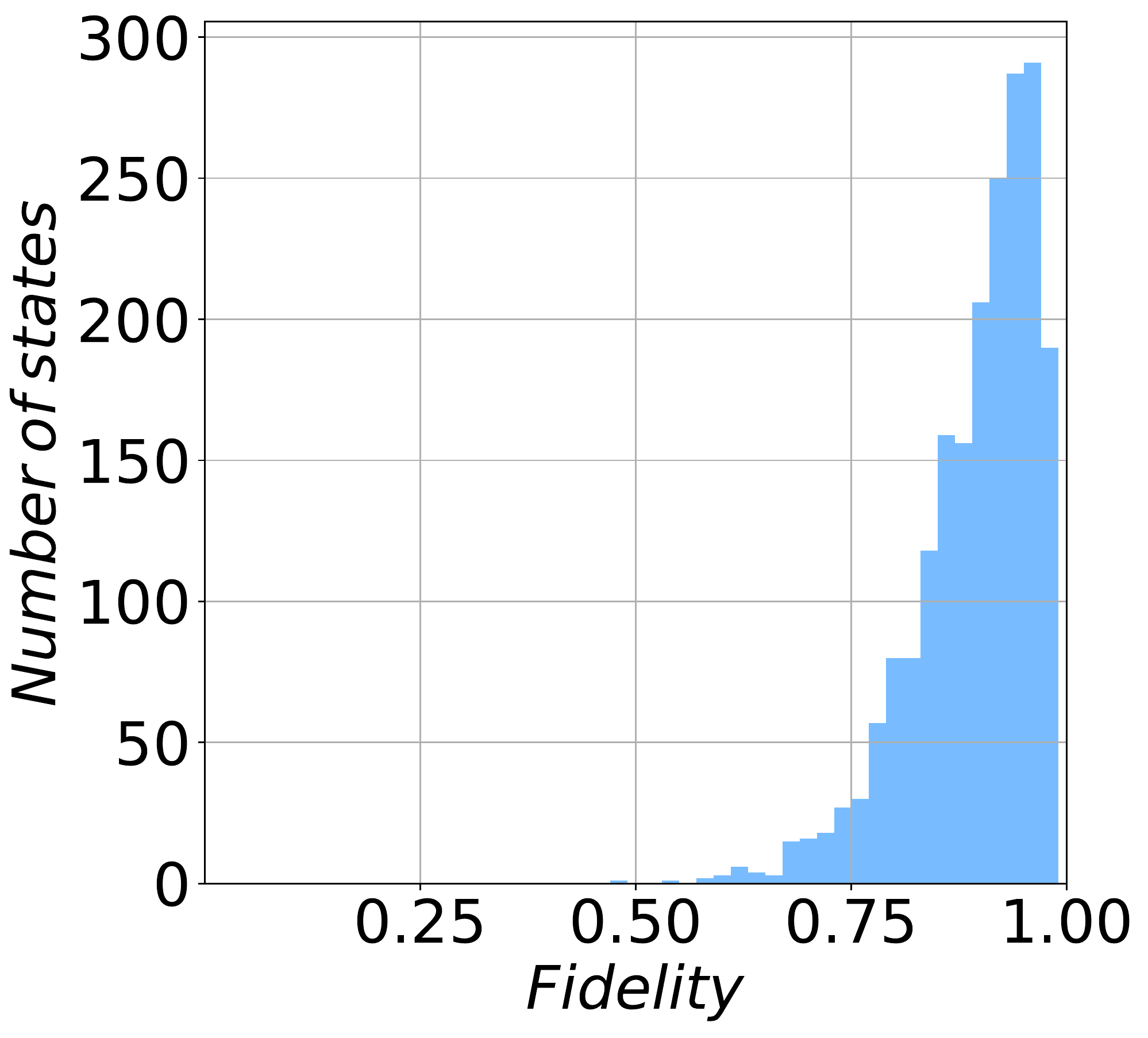}
    \label{fig:Fid_3-3_corr}}
\end{minipage}
\begin{minipage}[b]{0.48\columnwidth}
\centering
\subfloat[Inverted coefficients]{\includegraphics[width=0.98\textwidth]{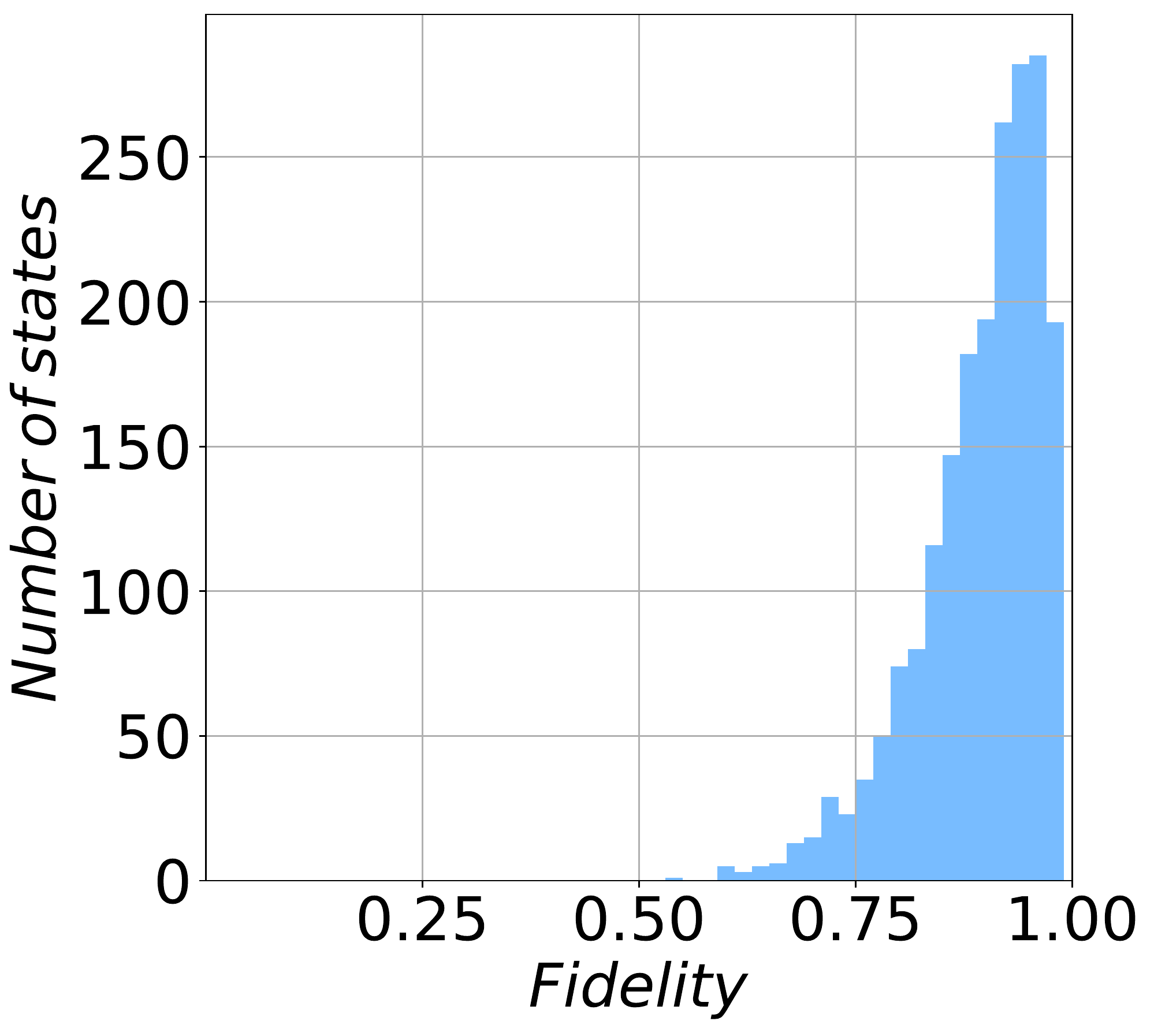}
    \label{fig:Fid_3-3_inv}}
\end{minipage}
\\
\centering
{Double Image Configuration}\\
\begin{minipage}[c]{0.48\columnwidth}
\subfloat[Correct coefficients]{\includegraphics[width=0.98\textwidth]{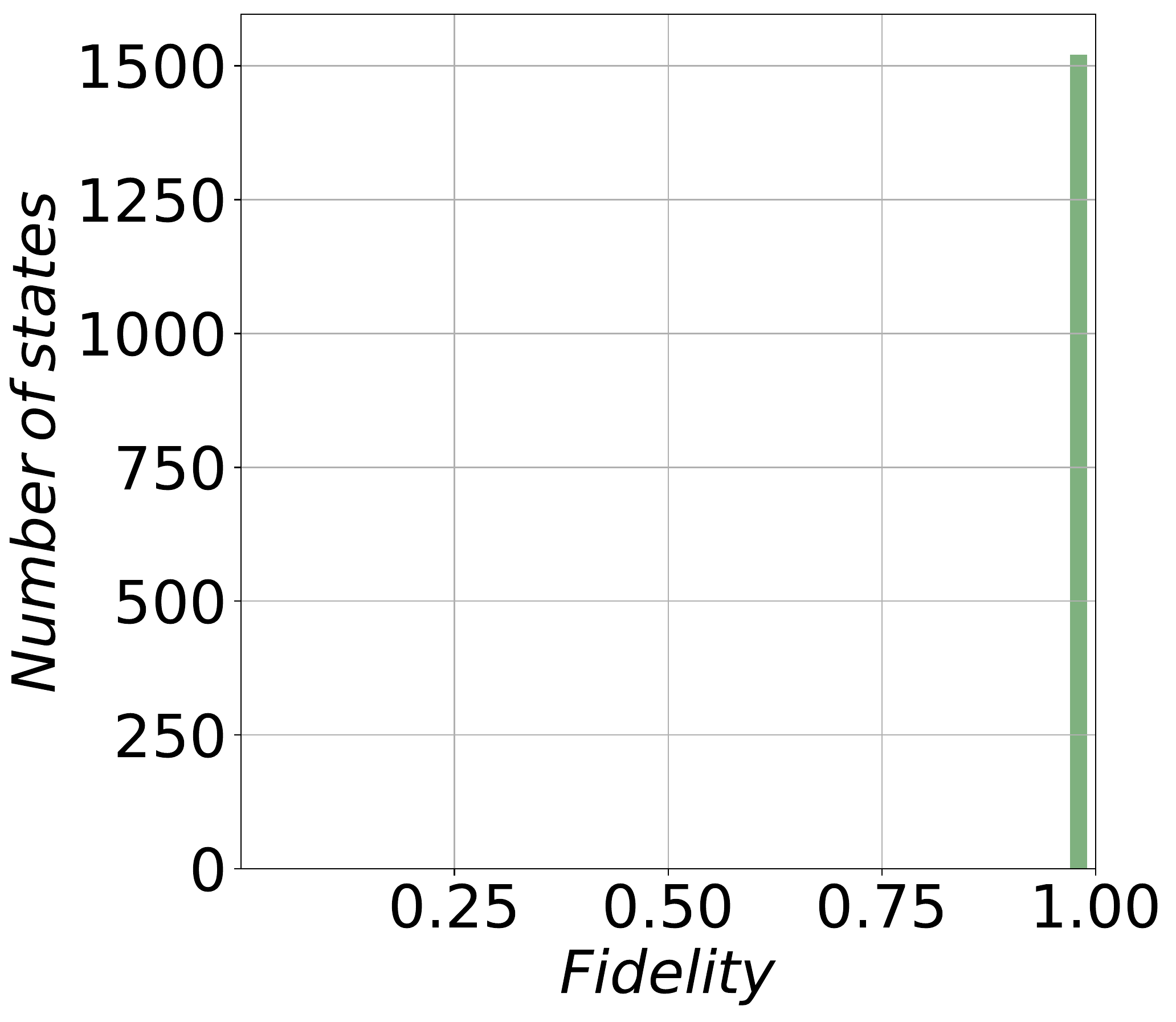}
    \label{fig:Fid_3-3_corr}}
\end{minipage}
\begin{minipage}[d]{0.48\columnwidth}
\centering
\subfloat[Inverted coefficients]{\includegraphics[width=0.98\textwidth]{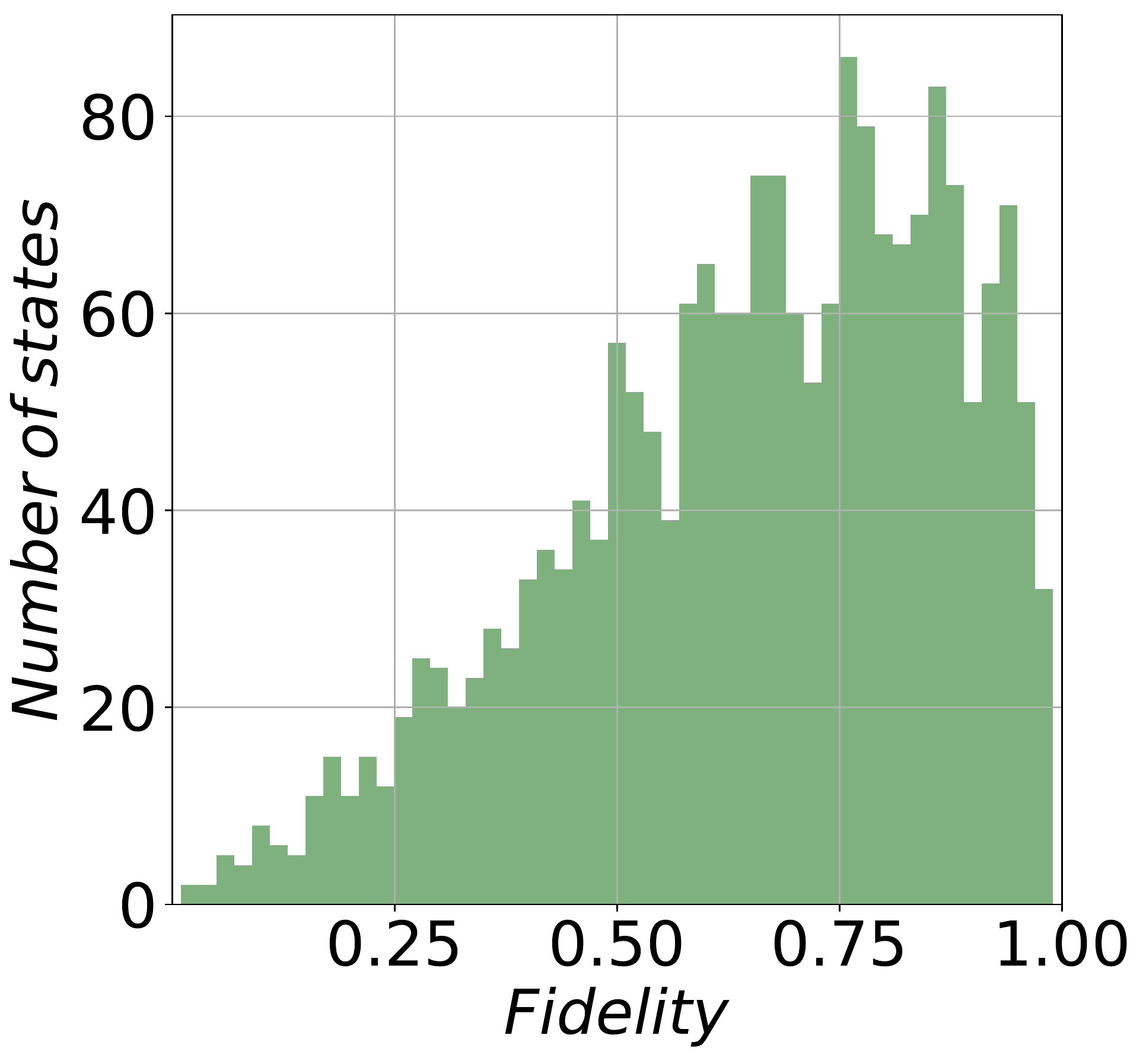}
    \label{fig:Fid_3-3_inv}}
\end{minipage}
\caption[.]{\textbf{Four-dimensional fidelities distribution.} Fidelity calculated on theoretical superpositions with $l \in \lbrace -3,-1,+1,+3 \rbrace$.
    The values of fidelity were calculated between the output of the regression with both the exact theoretical state (a) and the theoretical state after the inversion of the coefficients described by Eq.~\eqref{eq:symmetry} (b).
    The fidelity values are worse than those presented in Fig.~\ref{fig:Fid_1-1_corretta_inversa} ($F_a=0.904(2)$ and $F_b=0.904(2)$), but they still are similar, meaning that the system does not distinguish the two cases. Computing the same fidelity in the double image configuration, the mean value of graph in panel (c) is $\Bar{F}_c=1$, while the mean value of graph in panel (d) is $\Bar{F}_d=0.664(5)$.
    They are incompatible and thus the process has broken the symmetry. }\label{fig:Fid_3-3_corretta_inversa}
\end{figure}

\section{Observation of symmetry breaking}
\label{App_Simmetria}
As explained in the main text, OAM superpositions show a symmetry that makes impossible to distinguish some states through an intensity measurement because they present the same spatial profile for it. One aspect of particular interest is the behaviour of the regressor in the case it does not have access to the information required to break the symmetry, namely the image in which the value of the OAM is increased by one for each mode in the superposition. 	
In order to better understand this process, we began working on the simplest space i.e. with $l \in \lbrace -1,+1 \rbrace$, whose Bloch representation, the Bloch sphere, is well known and easy to analyze. 
We observed the results of the regression for both the case in which we break and don't break the symmetry and compared the positions of the regressed states on the Bloch sphere, these results are reported in Fig.~\ref{fig:PCA_1step_nosfera-tu} in the main text.
These results give us precise insight into what strategy the regression algorithm takes in order to mitigate the effect of the missing information. 
First of all, it is obvious that the states at the poles of the Bloch sphere (along the $z$ axis), which correspond to either $\ket{\psi}=\ket{-1}$ or $\ket{\psi}=\ket{+1}$, would be the ones on which regression would completely fail. 
Secondly, it can be seen that the effect on the Bloch sphere of the transformation described by Eq.~\eqref{eq:symmetry} is the reflection symmetry along the x-y plane. As a consequence, the points on the equator are invariant under this transformation and, for the same reasoning, it is not surprising that these points are correctly identified. 
Therefore, when the algorithm fails to break the symmetry, it tends to put the states near the equator in order to limit the errors in the state reconstruction. In fact, as shown in Fig. \ref{fig:PCA_1step_nosfera-tu}-b, as soon as we give the algorithm the necessary information, the states are placed in the right positions on the Bloch sphere.
To verify this behavior, in both single and double image configuration, we compared the regressed state with the expected theoretical state and the theoretical state on which Eq.~\eqref{eq:symmetry} has been applied.

In Fig.~\ref{fig:Fid_1-1_corr} and ~\ref{fig:Fid_1-1_inv} are shown the resulting distribution of the computed fidelities in the single image configuration and they are indistinguishable. 
Moreover, the relative mean fidelities are compatible within their statistical errors. 
Therefore, the regressor is not able to break the symmetry using a single image but only if two images are provided. Indeed, in Fig.~\ref{fig:Fid_1-1_corr_double} is reported the resulting distribution of the computed fidelity with the exact theoretical state and it is equal to $1$, while the distribution obtained comparing the regressed state with the theoretical state on which Eq.~\eqref{eq:symmetry} has been applied is lower than $80\%$ (see Fig.~\ref{fig:Fid_1-1_inv_double}). Therefore, the distributions are completely incompatible and the algorithm managed to break the symmetry.

\begin{figure}[!hb]
    \centering
    \includegraphics[width=1\columnwidth]{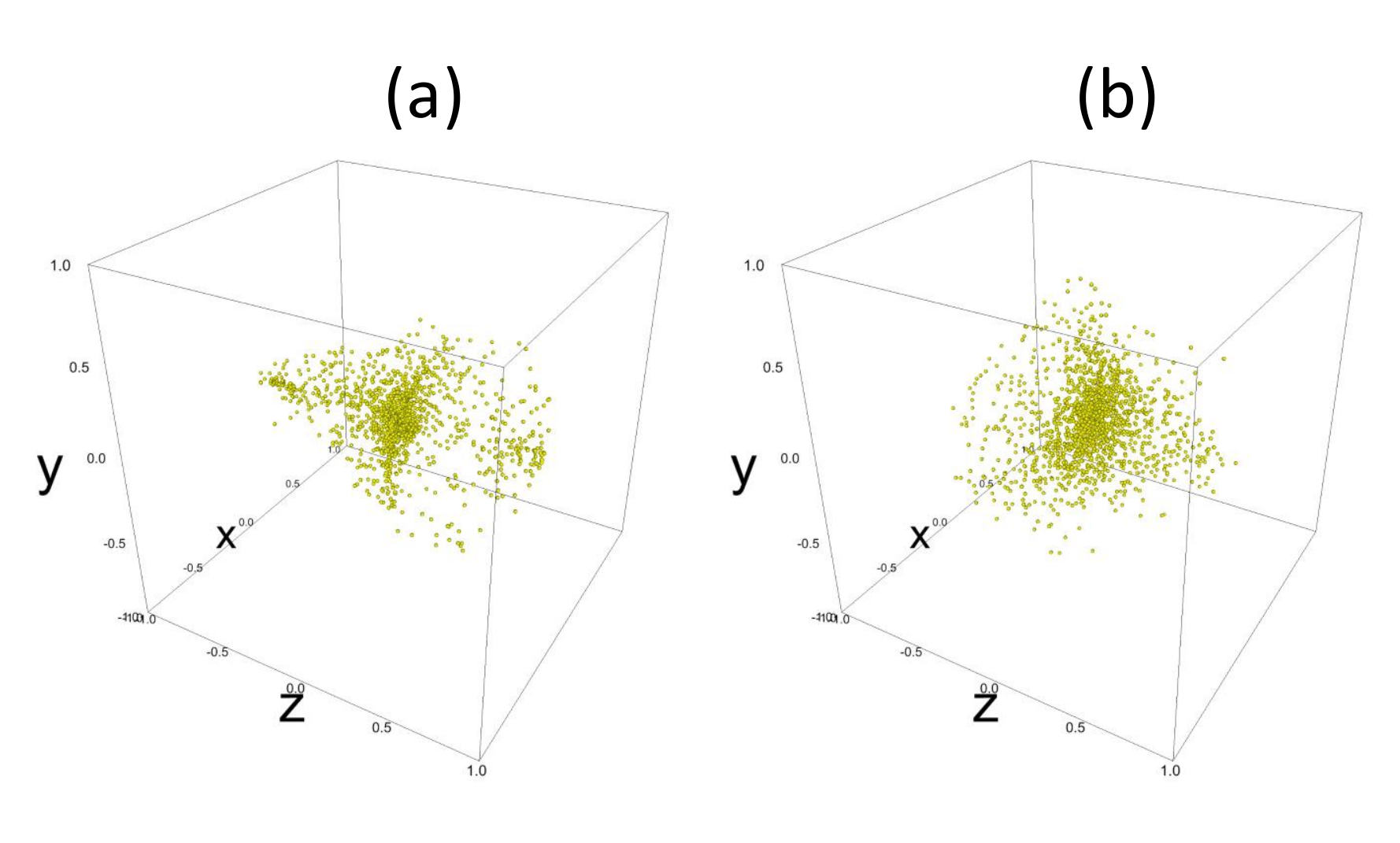}
    \caption{
    \textbf{Four-dimensional symmetry breaking.} Representation of the output of the regression algorithm on theoretical superpositions with $l \in \lbrace -3,-1,+1,+3 \rbrace$, after the projection on the nearest pure state. We compare the position of each state on the Bloch space obtained from the regressor using  one image (a) and two images (b) per each of them. The distribution of the states in this three dimensional space is quite different. As for the bi-dimensional case of Fig.~\ref{fig:PCA_1step_nosfera-tu},  this give an intuition on how the information stored in the image with augmented OAM values helps in obtain the correct states. This is explicitly shown in the fidelities distribution reported in Fig.~\ref{fig:Fid_3-3_corretta_inversa}. In each plot are shown only the first three dimensions of a fifteen dimensional space.} 
    \label{fig:PCA_3step_Aletta_O}
    \end{figure}
    
\;\\
\;\\
The same study was conducted for higher dimension superposition, in particular for $l \in \lbrace -3,-1,+1,+3 \rbrace$.
As expected, we observe that the two profiles still match and the two mean values are still compatible within their statistical errors (Fig.~\ref{fig:Fid_3-3_corretta_inversa}-a,b). On the other side, comparing the distributions obtained in the double image configuration a substantial difference can be observed. The comparison with the correct theoretical state results in a unitary fidelity, while computing the fidelity with the theoretical state on which Eq.~\eqref{eq:symmetry} has been applied, each state is reconstructed almost randomly and it results in a mean fidelity lower than $70\%$. (Fig.~\ref{fig:Fid_3-3_corretta_inversa}-c,d). This allows us to extend the same conclusion we had on the two dimensional case to this higher dimensional one. However, since the dimension of the Bloch vector space increases from three to fifteen, it is more complex to perform a comparison like that in Fig.~\ref{fig:PCA_1step_nosfera-tu}. 
Nevertheless, in Fig.~\ref{fig:PCA_3step_Aletta_O} it is shown an example similar to that shown before, obtained by projecting the results on the first three dimensions of the Bloch vector. 

The results reported in this section point out one more time how it is necessary to break the symmetry in order to correctly detect the OAM content of arbitrary high dimensional superpositions.

\begin{figure}[t!]
\centering
{(a) Linear Regressor}
\begin{minipage}[b]{1\columnwidth}
\subfloat{\includegraphics[width=0.98\textwidth]{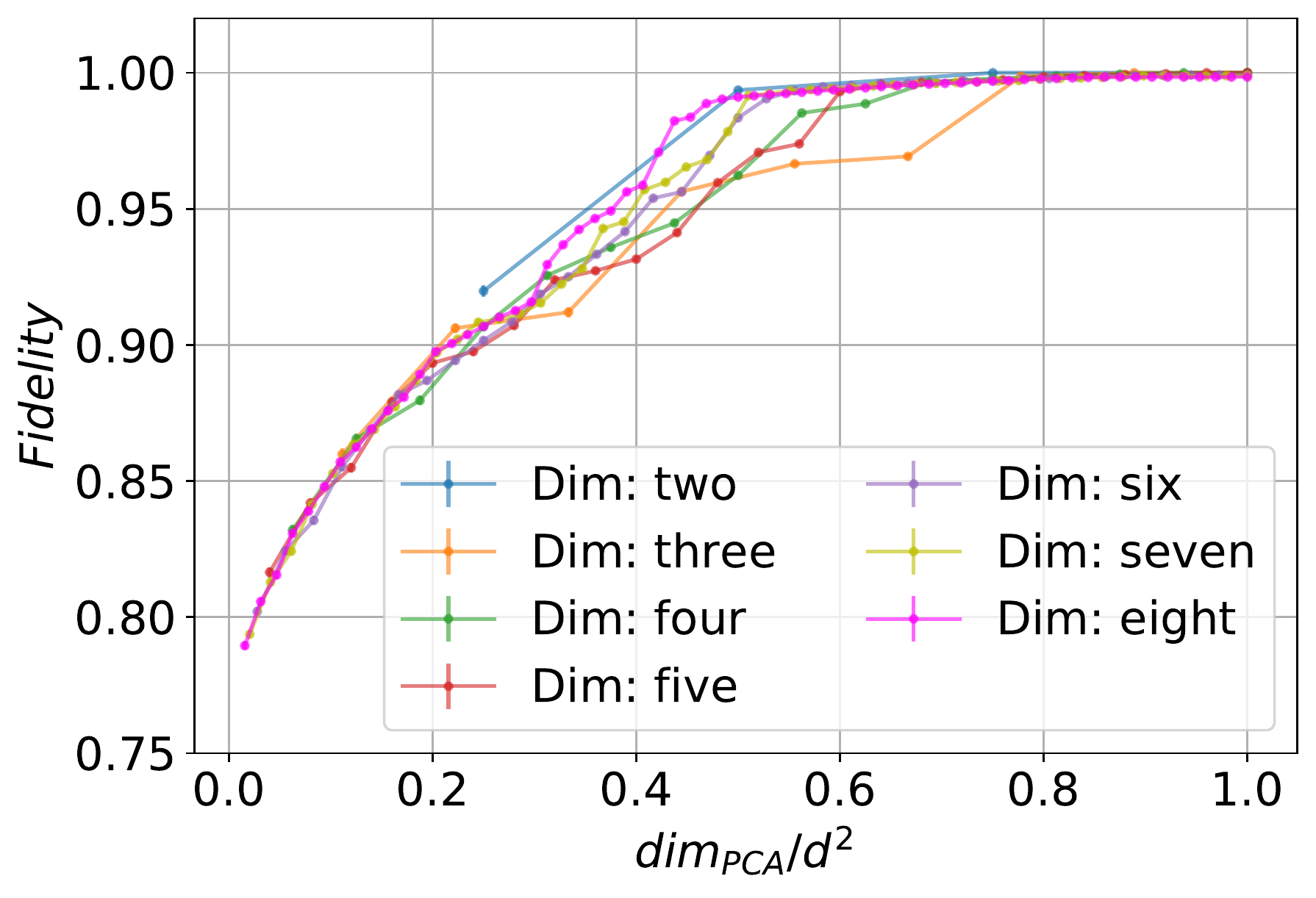}
    \label{fig:Fid_th_linear}}
\end{minipage}
\\
\centering
{(b) Extra Tree Regressor}
\begin{minipage}[b]{1\columnwidth}
\subfloat{\includegraphics[width=0.98\textwidth]{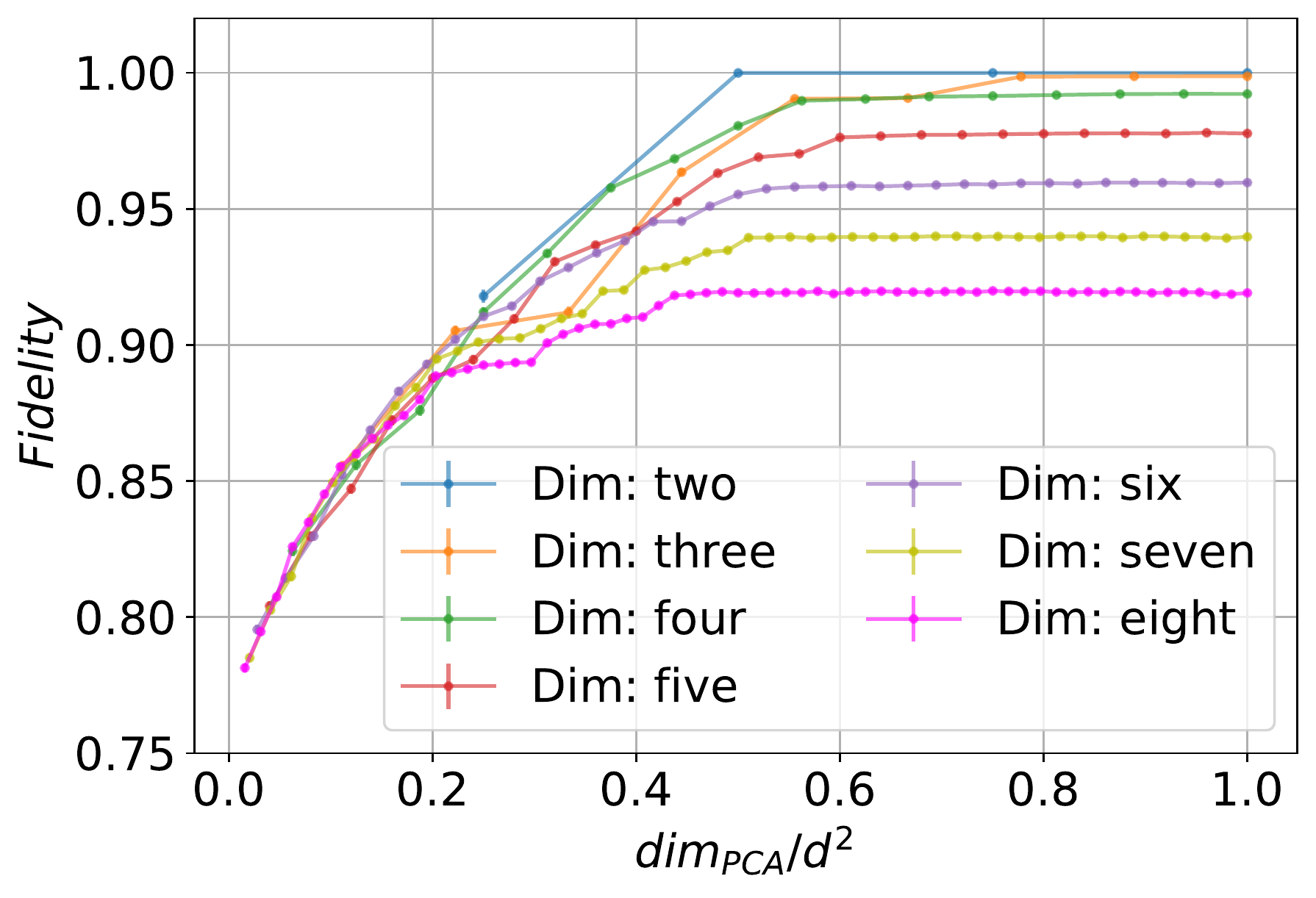}
    \label{fig:Fid_th_ETR}}
\end{minipage}
\caption[.]{\textbf{Comparison between regressor algorithms.} The plots show the results of the linear regressor (a) and of the extra tree regressor (b) when applied to theoretically simulated states with Hilbert space dimension up to 8. The number of PCA components given to the regressors is reported on the x axis, in order to consider various dimensions in a single plot we normalize it to the factor $d^2$. 
From the computed fidelities it can be seen that the linear regressor outperforms the ETR, reaching a value of the fidelity nearly equal to 1 for all the dimensions when we approach the value of $(d^2-1)$ PCA components, which is the dimension of the Bloch vector. Moreover, only the ETR presents a damping in the performances when the dimensions of the regressed states augment. All the fidelity values are obtained averaging over 2000 random states.}\label{fig:Fid_th_vs_PCA}
\end{figure}
\section{Comparison between  linear and nonlinear regression algorithms}
\label{Confronto}
In this section, we compare the performances of the linear regressor used in this work with those of a nonlinear regressor called extra tree regressor \cite{scikit-learn}. The latter is a predictor that exploits the structure of decision trees to associate each input with its correct label. The decision tree is composed of several nodes. In each of these, a condition on the input space determines which direction has to be followed and which of the children nodes will be the next. This process is repeated until the arrival at any leaf node without children which contains the value of the regressor output.

We compare the performances of the ETR with the linear regressor used in the work, by considering simulated OAM superposition states with a dimension of the Hilbert space $d \in [2,8]$. In particular, we generated 10000 random states, using 8000 of them as training set and the remaining as test set, and gave them as input to the PCA and thereafter to the two regressors. In all the cases under analysis, the double image approach described in the main text is used. 
In particular, we studied the fidelity behavior of the predicted states changing the number of PCA components given to the regressors. The resulting mean values of the fidelity, obtained averaging over the test set, are reported in Fig. \ref{fig:Fid_th_vs_PCA}. These showcase how the linear regressor outperforms the ETR for all the cases considered, in fact it reaches higher values for the fidelity not having the damping in the performances presented by ETR when the dimension increases. 
This effect is mostly caused by the purely interpolating action of the ETR model, meaning that for high dimensional spaces it usually concentrates the states towards the origin of the Bloch space. This generates additional error that increases with dimension of the space, making such model less suitable for high dimensions. 
We also note that the linear regressor obtains a value for the mean fidelities nearly equal to 1 when the number of PCA components is close to $(d^2-1)$, i.e. the dimension of the Bloch vector describing the state.
The obtained results highlight how, due to the linearity in the PCA action, an approach based on the use of a linear regressor is better suited to solve the task without the need of more complex approaches.


%

\end{document}